# Local geometry of electromagnetic fields and its role in molecular multipole transitions

*Nan Yang[1] and Adam E. Cohen[2,]*

[1]School of Engineering and Applied Sciences, Harvard University, Cambridge, MA 02138, USA

[2]Departments of Chemistry and Chemical Biology and Physics, Harvard University, 12 Oxford Street, Cambridge, MA 02138, USA

* To whom correspondence should be addressed. cohen@chemistry.harvard.edu

**Abstract**

Electromagnetic fields with complex spatial variation routinely arise in Nature. We study the response of a small molecule to monochromatic fields of arbitrary three-dimensional geometry. First, we consider the allowed configurations of the fields and field gradients at a single point in space. Many configurations cannot be generated from a single plane wave, regardless of polarization, but any allowed configuration can be generated by superposition of multiple plane waves. There is no local configuration of the fields and gradients that *requires* near-field effects. Second, we derive a set of local electromagnetic quantities, where each couples to a particular multipole transition. These quantities are small or zero in plane waves, but can be large in regions of certain superpositions of plane waves. Our findings provide a systematic framework for designing far-field and near-field experiments to drive multipole transitions. The proposed experiments provide information on molecular structure that is inaccessible to other spectroscopic techniques, and open the possibility for new types of optical control of molecules.

## 1. Introduction

Propagating plane waves comprise only a minute fraction of all solutions to Maxwell's Equations. Fields with sinusoidal time dependence but with more complex spatial variation routinely arise in the context of multiple-wave interference and optical near fields. These variously shaped fields can excite molecular multipole transitions that are tickled weakly—if at all—by far-field plane waves. Here we study the geometry of non-plane-wave electromagnetic fields, and the linear interaction of these fields with small molecules. For each type of multipole transition we present a local electromagnetic quantity that determines the strength of the coupling to that transition. We propose simple field configurations and spectroscopic techniques that selectively probe particular multipole transitions.

For most molecules, far-field plane waves couple weakly to molecular multipole transitions beyond electric dipole, and in some cases the coupling is identically zero. Weak coupling beyond electric dipole arises because the wavelength of light is typically much larger than molecular dimensions. Higher multipole transitions are excited more weakly than electric dipole by powers of ($a/\lambda$), where $a$ is the molecular size and $\lambda$ is the wavelength. Additionally, excitation of some multipole transitions by plane waves is identically zero due to the definite relations among electric fields, electric field gradients and magnetic fields that arise in all plane waves. These transitions vanish for plane waves regardless of the wavelength, polarization, or degree of orientation of the molecules.

Steeply varying fields are widespread in Nature. They play a particularly important role in intermolecular interactions because the electromagnetic field due to one molecule may be highly nonuniform over the extent of its neighbor. These field gradients couple to multipole transitions beyond electric dipole that are important in mediating intermolecular energy transfer[1], intermolecular forces[2], and chemical reactions[3]. To understand these processes[4, 5], one would like to explore the response of a test molecule to a variety of time-harmonic electric fields, magnetic fields, and field gradients.

Metallic nanostructures, photonic crystals, and metamaterials also generate local fields that are highly contorted. Surface-enhanced optical effects are well known for molecules near metal surfaces[6, 7], and include enhanced fluorescence[8], Raman scattering[9], two-photon excitation[10], and photochemistry[11]. These phenomena are typically interpreted in terms of enhancement of the electric field strength alone[6]; but the relative magnitude and direction of electric and magnetic fields and their gradients near a nanostructure need not correspond with their values in a plane wave, and can thereby violate far-field selection rules. For instance, it was recently predicted that magnetic nanostructures can dramatically enhance the rate of intersystem crossing in nearby molecules through interaction of magnetic gradients with a magnetic quadrupole transition in a radical pair[12]. Nanostructures can sculpt the fields to bring "forbidden" transitions to light.

Thus it is interesting to study the response of molecules to non-plane-wave electromagnetic fields. In section 2 we study the possible local geometries of monochromatic fields allowed by Maxwell's Equations. There are many valid field configurations that cannot be produced by a monochromatic plane wave. This distinction between the space of possible fields, and the space of plane wave fields is an important aspect of spectroscopy that has not received adequate attention.

We show that *any* valid configuration of electric and magnetic fields and field gradients can be created at discrete points in space by superposition of up to 32 linearly polarized monochromatic plane waves. Thus, while it may be convenient to use nanostructures or near-field optics to generate certain field configurations, these tools are not strictly necessary.

Next we ask: What attributes of the field should one calculate to determine the rate of excitation of a particular multipole transition? Our group previously considered this question in the context of randomly oriented chiral molecules and circular dichroism. We introduced a time-even pseudoscalar that measures the local handedness of the electric and magnetic fields:

$$C \equiv \frac{\varepsilon_0}{2} \vec{E} \cdot \nabla \times \vec{E} + \frac{1}{2\mu_0} \vec{B} \cdot \nabla \times \vec{B}. \tag{1}$$

This "Optical Chirality" determines the dissymmetry in the rate of excitation of a small chiral molecule[13].

In section 3 we study how other bilinear field objects couple to other kinds of multipole transitions. In section 3.2 we treat the coupling of electric field gradients to electric dipole-electric quadrupole (E1-E2) transitions. These transitions average to zero for randomly oriented molecules, regardless of the field geometry. Near an interface, however, molecules may have a uniaxial orientation, leading to several interface-selective signals for appropriately sculpted fields.

In Section 3.3 we also treat electric dipole-magnetic dipole (E1-M1) transitions. Upon orientational averaging, these transitions survive only for chiral molecules and chiral fields. For uniaxially oriented molecules, however, E1-M1 transitions combine with E1-E2 transitions to give an interface-selective signal for appropriately sculpted fields. Remarkably, many of these transitions are completely invisible to plane waves, but become visible in a standing wave.

In sections 3.4-3.6, we construct simple standing wave fields which selectively excite E1-E2 and E1-M1 transitions in localized regions of space. We propose a remarkably simple experiment to test these predictions.

In section 4 we consider magnetic circular dichroism (MCD) as a perturbation to the electric dipole transition. We derive the electromagnetic quantity that couples to MCD and propose an experiment in which a focused beam of *linearly* polarized light probes the same molecular quantities that are usually probed with circularly polarized light in MCD.

In 1964 Lipkin introduced ten conserved electromagnetic quantities that are quadratic in the fields, which he called the "Zilch".[14] He and subsequent workers failed to find any physical meaning for the Zilch.[15, 16] One of these, termed $Z^{000}$, is the same as the optical chirality we previously introduced. In section 5 we show that six of the remaining Lipkin terms are the EM quantities that couple to E1-M1 transitions, and the other three couple to molecules in DC magnetic fields. We present a protocol for generating an arbitrary number of conserved Lipkin-like quantities.

## 2. Allowable monochromatic fields

How much freedom do Maxwell's Equations give us to sculpt electromagnetic fields? A small molecule is only sensitive to local fields and field gradients, so we restrict attention to these aspects of the field about a fixed point in space, chosen to lie at the origin $\mathbf{r} = 0$. A related question is: can we generate any allowed field configuration using combinations of plane waves, or are there configurations that only arise in optical near-fields? We show, perhaps contrary to intuition, that *any* time-harmonic local field configuration can be generated by superposition of propagating plane waves.

The most general monochromatic electric field follows an elliptical trajectory, with arbitrary orientation, amplitude, ellipticity, and phase. It is mathematically convenient to describe this configuration by the real part of

$$\widetilde{\mathbf{E}} = \widetilde{\mathbf{E}}^{(0)} e^{-i\omega t}, \qquad (2)$$

where $\widetilde{\mathbf{E}}$ and $\widetilde{\mathbf{E}}^{(0)}$ are complex vectors. We use symbols without tilde to represent physical quantities, which are the real parts of the corresponding complex terms. A single propagating plane wave can generate any desired electric field ellipse. One simply chooses the polarization, amplitude, and direction of propagation of the wave.

Within the electric point-dipole approximation (PDA), the local electric field is the only quantity needed to describe the interaction of light with matter. In this case plane waves span the space of relevant fields, so the distinction between plane waves and other fields is moot. However, if one considers multipole transitions beyond electric dipole, then the magnetic field and the electric and magnetic field gradients become important. In this case we find allowable field configurations that cannot be produced by a single plane wave.

One can imagine a monochromatic field configuration that at a single point consists of an electric field ellipse and a magnetic field ellipse, each with arbitrary amplitude, ellipticity, orientation, and phase (Figure 1). This configuration is consistent with Maxwell's Equations, but unless **E** and **B** are always perpendicular, it does not occur in a plane wave. First we show by explicit construction how to generate this configuration by combination of multiple standing waves. Then we use an algebraic approach to show that any electric and magnetic field *gradients* allowed by Maxwell's Equations can be superimposed on these ellipses by addition of more plane waves.

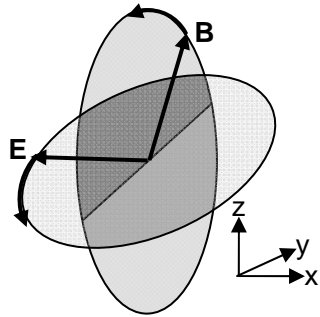

**Figure 1.** Arbitrary electric and magnetic ellipses at a single point in space. This configuration cannot occur in a plane wave, but can occur in a combination of two standing waves.

Consider a standing wave composed of counterpropagating plane waves, linearly polarized in the same plane. At the electric anti-nodes, the magnetic field is zero, and *vice versa*. By overlapping the electric anti-nodes of two such standing waves, with different linear polarizations, phases, and propagation axes, one creates an arbitrary electric field ellipse at a point where the magnetic field is zero. By combining the magnetic anti-nodes of two more standing waves in a similar manner, one creates an arbitrary magnetic ellipse where the electric field is zero. Finally, one superimposes the electric ellipse with the magnetic ellipse, to achieve the arbitrary field configuration shown in Figure 1. We describe the state of the field in Figure 1 by a six-element complex vector, $(\widetilde{\mathbf{E}}, \widetilde{\mathbf{B}})$.

Now we consider the electric and magnetic field gradients. The electric gradient $\nabla \tilde{\mathbf{E}}$ has nine components $\tilde{E}_{\beta\gamma}$, indicating the derivative of $\tilde{E}_\beta$ in the $\gamma$ direction. The magnetic gradient also has nine components. Once we have determined the vector $(\tilde{\mathbf{E}}, \tilde{\mathbf{B}})$, Maxwell's Equations set the divergence and curl of $\tilde{\mathbf{E}}$ and $\tilde{\mathbf{B}}$, leaving us with five independent components for $\tilde{E}_{\beta\gamma}$ and five for $\tilde{B}_{\beta\gamma}$. Thus we have 6 independent components for the fields, and 10 for the gradients. We describe the complete field geometry at a single point by a vector, $\tilde{\mathbf{Q}} \equiv (\tilde{\mathbf{E}}, \tilde{\mathbf{B}}, \nabla\tilde{\mathbf{E}}, \nabla\tilde{\mathbf{B}})$, with 16 independent complex components.

Now the question is: can any value of $\tilde{\mathbf{Q}}$ be created by superposition of linearly polarized plane waves? Every plane wave contributes to $\tilde{\mathbf{E}}$, $\tilde{\mathbf{B}}$, $\nabla\tilde{\mathbf{E}}$, and $\nabla\tilde{\mathbf{B}}$, and at a single point in space the wave can be represented as a vector in the space of $\tilde{\mathbf{Q}}$. If we can find a set of plane waves that span the space of $\tilde{\mathbf{Q}}$, then the problem is solved.

To construct a basis set for $\tilde{\mathbf{Q}}$, we restrict attention to linearly polarized, monochromatic plane waves, each with fixed wavevector and polarization, but variable amplitude. These are of the form:

$$\tilde{\mathbf{E}} = \mathbf{E}^{(0)} \exp(i\mathbf{k}\cdot\mathbf{r} - i\omega t)\exp(i\phi)$$

$$\tilde{\mathbf{B}} = \frac{1}{c}\hat{\mathbf{k}} \times \tilde{\mathbf{E}}$$

$$\tilde{E}_{\beta\gamma} = iE_\beta^{(0)} k_\gamma \exp(i\mathbf{k}\cdot\mathbf{r} - i\omega t)\exp(i\phi)$$

$$\tilde{B}_{\beta\gamma} = iB_\beta^{(0)} k_\gamma \exp(i\mathbf{k}\cdot\mathbf{r} - i\omega t)\exp(i\phi) \tag{3}$$

with $\mathbf{E}^{(0)}$ and $\mathbf{k}$ real. We have the constraints:

$$\mathbf{E}^{(0)} \cdot \mathbf{k} = 0, \tag{4}$$

$$|\mathbf{k}| = k_0. \tag{5}$$

For each plane wave we evaluate its contribution to $\tilde{\mathbf{Q}}$ at the origin. In Table 1 of Appendix 1 we list 16 monochromatic plane waves with $\phi = 0$ which span the real part of $\tilde{\mathbf{Q}}$. The same 16 waves with $\phi = \pi/2$ span the imaginary part of $\tilde{\mathbf{Q}}$.

Thus it is possible, using only plane waves, to achieve an arbitrary configuration of EM fields and gradients at a point in space. Nanostructures and near-field optics may offer practical advantages for the creation of these fields, but these tools are not strictly necessary. We now consider the response of a molecule to a field with an arbitrary value of $\tilde{\mathbf{Q}}$.

### 3. Molecular Multipole Transitions

**3.1 Induced Oscillating Molecular Multipoles.** A molecule subjected to a time-harmonic EM field develops time-harmonic charge and current distributions, which may be described by a multipole expansion:[17]

$$\begin{aligned}
\tilde{\mu}_\alpha &= \tilde{\alpha}_{\alpha\beta} \tilde{E}_\beta + \frac{1}{3} \tilde{A}_{\alpha\beta\gamma} \tilde{E}_{\beta\gamma} + \tilde{G}_{\alpha\beta} \tilde{B}_\beta \\
\tilde{\theta}_{\alpha\beta} &= \tilde{a}_{\gamma\alpha\beta} \tilde{E}_\gamma + ... \\
\tilde{m}_\alpha &= \tilde{g}_{\alpha\beta} \tilde{E}_\beta + ...
\end{aligned} \tag{6}$$

where α, β, γ are Cartesian indices, and $\tilde{\mu}$, $\tilde{\theta}$, $\tilde{m}$ are the oscillating electric dipole, electric quadrupole and magnetic dipole, respectively. The quantity $\tilde{\alpha}_{\alpha\beta}$ is the dynamic electric dipole polarizability, $\tilde{A}_{\alpha\beta\gamma}$ and $\tilde{a}_{\gamma\alpha\beta}$ are the mixed electric dipole-quadrupole polarizabilities, and $\tilde{G}_{\alpha\beta}$ and $\tilde{g}_{\alpha\beta}$ are the mixed electric-magnetic dipole analogues. We use the Einstein summation convention in which repeated Cartesian indices are summed.

The transition matrix elements and the lineshape functions are each, in general, complex. The polarizability tensors such as $\tilde{A}_{\alpha\beta\gamma}$ and $\tilde{G}_{\alpha\beta}$ are products of matrix elements and lineshapes, so these are generally complex too. In the absence of a DC magnetic field, the dynamics respect time-reversal symmetry, and all molecular eigenfunctions can be chosen to be real[18]. In this case electric dipole and electric quadrupole transition matrix elements are purely real, and magnetic dipole transition matrix elements are purely imaginary. From the perturbation theory expressions for the molecular response tensors one can show that:

$$\begin{aligned}\tilde{A}_{\alpha\beta\gamma} &= \tilde{a}_{\gamma\alpha\beta} = A'_{\alpha\beta\gamma} + iA''_{\alpha\beta\gamma} \\ \tilde{G}_{\alpha\beta} &= -\tilde{g}_{\alpha\beta} = G'_{\alpha\beta} + iG''_{\alpha\beta}\end{aligned} \quad (7)$$

where $A'_{\alpha\beta\gamma}$, $A''_{\alpha\beta\gamma}$, $G'_{\alpha\beta}$ and $G''_{\alpha\beta}$ are real frequency-dependent functions. Note that we use a different notation from Barron[17] because we are restricting attention to conditions where the molecular eigenfunctions can be chosen to be real.

The induced multipoles absorb energy from the fields at an average rate

$$\Gamma = \left\langle E_\alpha \dot{\mu}_\alpha + \frac{1}{3} E_{\alpha\beta} \dot{\theta}_{\alpha\beta} + B_\alpha \dot{m}_\alpha + ... \right\rangle_t, \quad (8)$$

where $\langle \ \rangle_t$ indicates an average over time. Substitution of Eq. 6666 into Eq. 8 leads to an expression for the rate of absorption in terms of molecular properties and local EM fields. This expression contains many terms, each of which is a product of a molecular tensor component and two field quantities.

Each term is responsible for a spectroscopic observable. We classify these observables by the field components that contribute. For instance, if we express the electric field as in Eq. 2, then the first term in the expansion is

$$\Gamma_{E1-E1} = \omega \alpha''_{\alpha\beta} E^{(0)}_\alpha E^{(0)}_\beta, \tag{9}$$

which is responsible for the pure electric dipole absorption which usually dominates. We neglect terms of order M2 and higher, because these are negligibly small for most molecules.

In non-plane wave geometries, the relative strengths of the field quantities can be adjusted to enhance the contribution of particular molecular terms. This flexibility enables enhanced spectroscopy with sculpted fields, and in some cases gives rise to observables not probed by plane waves.

**3.2 Electric dipole – electric quadrupole (E1-E2) excitation.** Two effects contribute to E1-E2 transitions: (1) interaction of the electric field with dipole moments induced by the field gradient and (2) interaction of the field gradient with quadrupole moments induced by the electric field. We now treat each in turn.

The complex fields at the origin $\tilde{E}_\beta = \tilde{E}^{(0)}_\beta e^{-i\omega t}$, $\tilde{E}_{\beta\gamma} = \tilde{E}^{(0)}_{\beta\gamma} e^{-i\omega t}$, can be expressed as

$$\begin{aligned}\tilde{E}_\beta &= E_\beta + \frac{i}{\omega}\dot{E}_\beta \\ \tilde{E}_{\beta\gamma} &= E_{\beta\gamma} + \frac{i}{\omega}\dot{E}_{\beta\gamma}\end{aligned} \tag{10}$$

where $E_\beta$ and $E_{\beta\gamma}$ are the real oscillating field components. The electric dipole moment induced by an electric field gradient is given by combining Eqs. 666 and 10:

$$\tilde{\mu}_\alpha = \frac{1}{3}(A'_{\alpha\beta\gamma} + iA''_{\alpha\beta\gamma})\left(E_{\beta\gamma} + \frac{i}{\omega}\dot{E}_{\beta\gamma}\right). \tag{11}$$

The time-dependent absorption rate due to this interaction is:

$$E_\alpha \dot{\mu}_\alpha = \frac{1}{3}A'_{\alpha\beta\gamma} E_\alpha \dot{E}_{\beta\gamma} + \frac{\omega}{3}A''_{\alpha\beta\gamma} E_\alpha E_{\beta\gamma} \tag{12}$$

where $\mu_\alpha = \text{Re}(\tilde{\mu}_\alpha)$ is the real component of the complex electric dipole moment; and we used the fact that for time-harmonic fields, $\ddot{E}_{\beta\gamma} = -\omega^2 E_{\beta\gamma}$.

Similarly, the molecular electric quadrupole moment induced by the electric field is:

$$\tilde{\theta}_{\beta\gamma} = (A'_{\alpha\beta\gamma} + iA''_{\alpha\beta\gamma})\left(E_\alpha + \frac{i}{\omega}\dot{E}_\alpha\right), \tag{13}$$

where we have used the fact that $\tilde{A}_{\alpha\beta\gamma} = \tilde{a}_{\gamma\alpha\beta}$. The time-dependent absorption rate due to this interaction is

$$\frac{1}{3}E_{\beta\gamma}\dot{\theta}_{\beta\gamma} = \frac{1}{3}A'_{\alpha\beta\gamma}\dot{E}_\alpha E_{\beta\gamma} + \frac{\omega}{3}A''_{\alpha\beta\gamma} E_\alpha E_{\beta\gamma}. \tag{14}$$

The total absorption rate is the sum of Eqs. 1212 and 14. The electromagnetic quantity that couples to $A''_{\alpha\beta\gamma}$ is:

$$\frac{2\omega}{3}\langle E_\alpha E_{\beta\gamma}\rangle_t, \tag{15}$$

and the one that couples to $A'_{\alpha\beta\gamma}$ is:

$$\frac{1}{3}\langle E_\alpha \dot{E}_{\beta\gamma} + \dot{E}_\alpha E_{\beta\gamma}\rangle_t = \frac{1}{3}\left\langle \frac{\partial}{\partial t}(E_\alpha E_{\beta\gamma})\right\rangle_t, \tag{16}$$

which vanishes. Thus the rate of the E1-E2 transition is:

$$\Gamma_{E1-E2} = \frac{2\omega}{3} A''_{\alpha\beta\gamma} \langle E_\alpha E_{\beta\gamma}\rangle_t. \tag{17}$$

Terms that couple to $A''_{\alpha\beta\gamma}$ with all three indices distinct give rise to optical activity of oriented molecules[19]. The average of the third rank tensor $\tilde{A}_{\alpha\beta\gamma}$ over all molecular orientations is:

$$\langle \tilde{A}_{\alpha\beta\gamma}\rangle_\Omega = -\frac{1}{6}(\varepsilon_{\alpha\beta\gamma}\tilde{A}_{\alpha\beta\gamma})\vec{\varepsilon}, \tag{18}$$

where $\varepsilon_{\alpha\beta\gamma}$ is the third rank isotropic tensor. The tensor components of $\tilde{A}$ are related by $\tilde{A}_{\alpha\beta\gamma} = \tilde{A}_{\alpha\gamma\beta}$,[17] whence $\varepsilon_{\alpha\beta\gamma}\tilde{A}_{\alpha\beta\gamma} = 0$. Therefore the E1-E2 absorption vanishes for unoriented molecules, regardless of the geometry of the EM field.

**3.3 Electric Dipole - Magnetic Dipole (E1-M1) Transition.** Two effects also contribute to E1-M1 transitions: (1) interaction of the electric field with electric dipole moments induced by the magnetic field and (2) interaction of the magnetic field with magnetic dipole moments induced by the electric field. We now treat each in turn.

The magnetic field induces an electric dipole moment according to $\tilde{\mu}_\alpha = \tilde{G}_{\alpha\beta} \tilde{B}_\beta$. By following the same algebra used to derive Eq. 1212, we find that this term contributes a time-dependent absorption rate

$$E_\alpha \dot{\mu}_\alpha = G'_{\alpha\beta} E_\alpha \dot{B}_\beta + \omega G''_{\alpha\beta} E_\alpha B_\beta. \tag{19}$$

The electric field induces a magnetic dipole moment according to $\tilde{m}_\beta = \tilde{g}_{\beta\alpha} \tilde{E}_\alpha$. The absorption rate due to this interaction is

$$B_\beta \dot{m}_\beta = -G'_{\alpha\beta} \dot{E}_\alpha B_\beta - \omega G''_{\alpha\beta} E_\alpha B_\beta, \tag{20}$$

where we have used the fact that $\tilde{G}_{\alpha\beta} = -\tilde{g}_{\alpha\beta}$. The rates in Eqs. 19 and 20 are summed to give the total rate of E1-M1 absorption

$$\Gamma_{E1-M1} = G'_{\alpha\beta} \left\langle E_\alpha \dot{B}_\beta - \dot{E}_\alpha B_\beta \right\rangle_t. \tag{21}$$

Thus the quantity $\left\langle E_\alpha \dot{B}_\beta - \dot{E}_\alpha B_\beta \right\rangle_t$ determines the rate at which E1-M1 transitions are excited in oriented molecules.

In the case of unoriented molecules, we take the isotropic average and use $\left\langle G'_{\alpha\beta} \right\rangle_\Omega = \frac{1}{3}(G'_{\alpha\alpha})\mathbf{I}$, where $\mathbf{I}$ is the 3-by-3 identity matrix. This quantity is only nonzero for chiral molecules. Application of Maxwell's Equations in free space to the quantity $\left\langle E_\alpha \dot{B}_\alpha - \dot{E}_\alpha B_\alpha \right\rangle_t$ shows that this object is proportional to Optical Chirality, and is responsible for chiral dissymmetry in the excitation of isotropic molecules.

Higher multipole transitions such as E1-M2, E2-M1, and M1-M1 can be calculated in similar manner. However the molecular response tensors for these transitions tend to be very small, so we deem experimental observation unlikely in the near term.

**3.4. Multipole transitions in linearly polarized standing waves.** Here we demonstrate that a linearly polarized standing wave excites molecular transitions that are invisible to traveling waves. These "achiral multipole transitions" are only detectable for molecules that are uniaxially oriented, and lie within a plane of sub-wavelength thickness. This geometry is common at liquid interfaces, so we propose that achiral multipole transitions can be used for surface-sensitive spectroscopy.

Consider a standing wave composed of two waves counter-propagating along $z$ and polarized along $x$. We allow the two waves to have different amplitudes, $E_0$ and $E_0'$.

$$\begin{aligned}\tilde{\mathbf{E}}(z,t) &= E_0 \hat{\mathbf{e}}_x \exp(i(kz - \omega t)) - E_0' \hat{\mathbf{e}}_x \exp(i(-kz - \omega t)) \\ \tilde{\mathbf{B}}(z,t) &= \frac{E_0}{c} \hat{\mathbf{e}}_y \exp(i(kz - \omega t)) + \frac{E_0'}{c} \hat{\mathbf{e}}_y \exp(i(-kz - \omega t))\end{aligned} \quad (22)$$

The only contribution to E1-E2 absorption is from $E_x E_{xz}$, and the only contribution to E1-M1 absorption is from $E_x \dot{B}_y - \dot{E}_x B_y$. The relevant field objects are:

$$\text{E1-E1:} \quad \omega \langle E_x^2 \rangle_t = \frac{\omega}{2}\left[E_0^2 + E_0'^2 - 2E_0 E_0' \cos(2kz)\right] \quad (23)$$

$$\text{Achiral E1-E2:} \quad \frac{2\omega}{3}\langle E_x E_{xz} \rangle_t = \frac{\omega k}{3} E_0 E_0' \sin(2kz) \quad (24)$$

$$\text{Achiral E1-M1:} \quad \langle E_x \dot{B}_y - \dot{E}_x B_y \rangle_t = -k E_0 E_0' \sin(2kz). \quad (25)$$

Eq. 23 determines the spatial dependence of E1-E1 absorption, which is the usual quantity measured in a standing wave. Eqs. 24 and 25 determine the rates of achiral multipole transitions, and have the remarkable property that they vanish for a plane wave (setting $E_0$ or $E_0'$ to zero). Achiral multipole transitions are invisible to a single linearly polarized plane wave. We show below that these transitions are also invisible to circularly polarized light, supporting our contention that standing waves can see transitions that plane waves cannot.

Substitution of Eqs. 24 and 25 into the overall rate of absorption gives

$$\Gamma = \Gamma_{E1-E1} + kE_0 E_0' \sin(2kz)\left(\frac{\omega A_{xxz}''}{3} - G_{xy}'\right). \tag{26}$$

The $\sin(2kz)$ dependence of the achiral multipole transitions combines with the $\cos(2kz)$ dependence of the E1-E1 transition to create a slight phase shift in the standing wave absorption pattern relative to what would be expected for pure electric dipole absorption.

The consequences of Eq. 26 vary depending on the degree of orientational order of the molecules. Eq. 26 applies to perfectly oriented molecules, such as might be found in a crystal. Under isotropic averaging, as in a liquid, $\langle \tilde{A}_{\alpha\beta\gamma} \rangle_\Omega = 0$, and the off-diagonal components of $\tilde{G}_{\alpha\beta}$ average to zero, even for chiral molecules. Thus these transitions are undetectable in bulk liquid. Near a liquid interface, however, molecules may adopt a uniaxial orientation. A DC electric field can orient molecules too. If the orientation axis coincides with the optical $z$-axis, then we average Eq. 26 about $z$, to get:

$$\Gamma = \Gamma_{E1-E1} + \frac{1}{2}kE_0 E_0' \sin(2kz)\left( \frac{\omega}{3}(A''_{xxz} + A''_{yyz}) - G'_{xy} + G'_{yx} \right). \qquad (27)$$

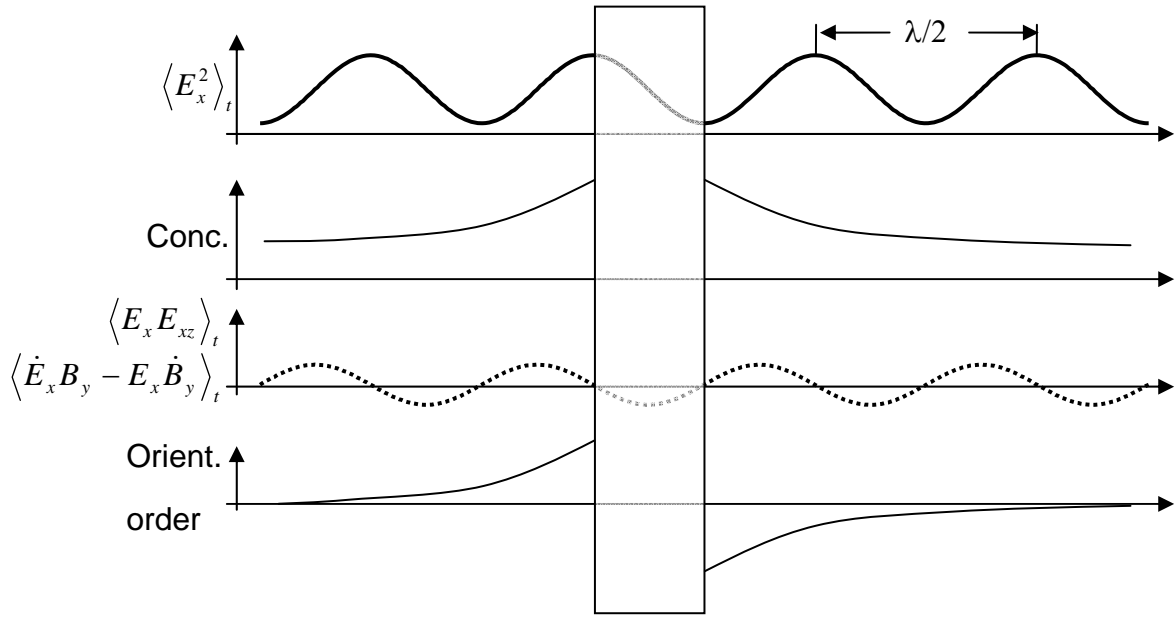

**Figure 2.** Schematic of an experiment to probe achiral E1-E2 and E1-M1 transitions in a liquid. A transparent barrier of thickness $(n/2 + ¼)\lambda$ separates two compartments of liquid and is exposed to a standing wave (for clarity we draw a slab of thickness $\lambda/4$, though this is mechanically implausible). The standing wave is moved perpendicular to the plane of the slab. Molecules may have larger (or smaller) concentration near the interface than in bulk, and may have orientational order near the interface. Only E1-E2 and M1-M2 excitation of the surface-oriented molecules oscillates with translation of the standing wave; the total rate of E1-E1 excitation does not vary.

**3.5 Experiment to detect achiral multipole transitions.** We propose a conceptually simple experiment to probe the achiral multipole transitions of molecules at a liquid interface. A transparent slab of thickness $(n/2 + ¼)\lambda$, where $n$ is an integer, is immersed in a liquid. The liquid molecules near the surface are oriented by interaction with the slab, so the orientation is in the opposite sense on the two faces. A standing wave is generated perpendicular to the slab. The absorption is monitored, for instance by fluorescence, as the standing wave is translated along the optic axis (Figure 2).

The total fluorescence from the bulk is independent of the position of the standing wave. If there is an enrichment or depletion of molecules near the surfaces of the slab, the periodic modulation in the E1-E1 fluorescence from one face is cancelled by an out-of-phase modulation in the E1-E1 fluorescence from the opposite face (a consequence of the choice of slab thickness). But if the molecules are oriented by interaction with the faces of the slab, then both the orientation and the direction of the electric field gradient are opposite on the two faces. The achiral multipole transitions from the two phases thereby have *in phase* periodic modulation. One records the total fluorescence as a function of time, and the amplitude of the oscillatory component indicates the strength of these transitions. One can also conceive a variant of this experiment in which the liquid forms a thin film in a channel of thickness $(n/2 + ¼)\lambda$.

**3.6. Multipole transitions in circularly polarized standing waves.** Standing waves with circular polarization also excite certain multipole transitions with greater selectivity than is obtained from circularly polarized plane waves. Consider a standing wave consisting of counterpropagating left- and right-circularly polarized waves, with possibly different amplitudes:

$$\widetilde{\mathbf{E}}(z,t) = \frac{E_0}{\sqrt{2}}\left(\hat{\mathbf{e}}_x + i\hat{\mathbf{e}}_y\right)\exp(i(kz - \omega t)) - \frac{E_0'}{\sqrt{2}}\left(\hat{\mathbf{e}}_x + i\hat{\mathbf{e}}_y\right)\exp(i(-kz - \omega t)).$$

$$\tilde{\mathbf{B}}(z,t) = \frac{E_0}{c\sqrt{2}}(\hat{\mathbf{e}}_y - i\hat{\mathbf{e}}_x)\exp(i(kz-\omega t)) + \frac{E_0'}{c\sqrt{2}}(\hat{\mathbf{e}}_y - i\hat{\mathbf{e}}_x)\exp(i(-kz-\omega t)). \tag{28}$$

We call this the $\sigma^+\sigma^-$ configuration. The standing wave generates the following EM quantities which couple to multipole transitions:

$$\text{E1-E1:} \qquad \omega\langle|\mathbf{E}|^2\rangle_t = \frac{\omega}{2}\left[E_0^2 + E_0'^2 - 2E_0 E_0' \cos(2kz)\right] \tag{29}$$

$$\text{Chiral E1-E2:} \qquad \frac{2\omega}{3}<E_x E_{yz}>_t = -\frac{2\omega}{3}<E_y E_{xz}>_t = -\frac{\omega k}{6}(E_0^2 - E_0'^2) \tag{30}$$

$$\text{Achiral E1-E2:} \qquad \frac{2\omega}{3}<E_x E_{xz}>_t = \frac{2\omega}{3}<E_y E_{yz}>_t = \frac{\omega k}{3}\sin(2kz)E_0 E_0' \tag{31}$$

$$\text{Achiral E1-M1:} \qquad <E_x\dot{B}_y - \dot{E}_x B_y>_t = -<E_y\dot{B}_x - \dot{E}_y B_x> = -kE_0 E_0' \sin(2kz) \tag{32}$$

$$\text{Chiral E1-M1:} \qquad <E_x\dot{B}_x - \dot{E}_x B_x>_t = <E_y\dot{B}_y - \dot{E}_y B_y>_t = -\frac{(E_0^2 - E_0'^2)\omega}{2c}. \tag{33}$$

As we found with linear polarization, the achiral E1-E2 and E1-M1 transitions are not excited by a single plane wave (setting $E_0$ or $E_0'$ to zero), but are only found in the standing wave. These transitions are truly invisible to plane-wave spectroscopy.

We refer to terms containing $E_x E_{yz}$ and $E_y E_{xz}$ as chiral, because these are responsible for optical activity of oriented chiral molecules. Our group previously showed that there exist configurations of the EM field in which the ratio of optical chirality to electric energy density is enhanced relative to the ratio found in circularly polarized light (CPL)[13]. It is then interesting to ask whether the same fields show an enhancement in the relative strength of the chiral E1-E2 terms.

The chiral selectivity of a light-matter interaction is described by the dissymmetry factor, defined as the fractional difference in rate of excitation between two mirror-image configurations of the fields (or the molecule).

$$g \equiv \frac{2(\Gamma^+ - \Gamma^-)}{(\Gamma^+ + \Gamma^-)}. \tag{34}$$

We previously considered isotropic chiral molecules for which the only chiral transition is E1-M1. We found that in a near-node of a $\sigma^+\sigma^-$ standing wave,

$$g_{max} = g_{CPL} \frac{E_0 + E_0'}{E_0 - E_0'}. \tag{35}$$

As $E_0 \rightarrow E_0'$, this enhancement factor can become very large.

Comparison of Eqs. 30 and 33 shows that both equations have the same dependence on $E_0$ and $E_0'$ and that both equations are independent of position. Thus the dissymmetry factor for E1-E2 transitions in oriented chiral molecules in a $\sigma^+\sigma^-$ standing wave undergoes the same enhancement as for E1-M1 transitions in unoriented molecules. In this regard, the fields in regions near the electric minima are truly "superchiral" for the $\sigma^+\sigma^-$ configuration.

### 4. Magnetic Circular Dichroism

Just as the optical chirality, $C$ determines the local strength of circular dichroism, we expect another EM quantity, $\Xi$, to determine the local strength of magnetic CD (MCD). While $C$ is a time-even pseudo-scalar, we expect $\Xi$ to be a time-odd axial vector.

MCD arises through a magnetic modification to the electric dipole polarizability tensor[4,20]. In the presence of a DC magnetic field, the electric field-induced dipole moment becomes:

$$\tilde{\mu}_\alpha = \tilde{\alpha}_{\alpha\beta\gamma} \tilde{E}_\beta B_\gamma$$
$$= (\alpha'_{\alpha\beta\gamma} + i\alpha''_{\alpha\beta\gamma})(E_\beta + \frac{i}{\omega}\dot{E}_\beta)B_\gamma \quad (36)$$

where $\tilde{\alpha}_{\alpha\beta\gamma}$ is the perturbation to the electric polarizability tensor by the static external field $B_\gamma$. The term proportional to $\alpha'_{\alpha\beta\gamma}$ is responsible for MCD [17]. The rate of absorption is

$$E_\alpha \dot{\mu}_\alpha = \alpha'_{\alpha\beta\gamma} E_\alpha \dot{E}_\beta B_\gamma. \quad (37)$$

As with regular absorption, MCD arises through a purely E1-E1 transition, in the sense that only the oscillating electric fields of the incident light need to be considered. The static magnetic field cannot contribute to absorption. However, the tensor $\tilde{\alpha}_{\alpha\beta\gamma}$ contains an antisymmetric component with respect to exchange of the first two indices, in contrast to the usual polarizability $\tilde{\alpha}_{\alpha\beta}$ which is symmetric. This antisymmetry explains why the real part of $\tilde{\alpha}_{\alpha\beta\gamma}$ determines MCD, while the imaginary part of $\tilde{\alpha}_{\alpha\beta}$ determines regular absorption.

If we consider randomly oriented molecules, then the relevant isotropically averaged field quantity is:

$$\varepsilon_{\alpha\beta\gamma} E_\alpha \dot{E}_\beta B_\gamma = (\mathbf{E} \times \dot{\mathbf{E}}) \cdot \mathbf{B}. \quad (38)$$

Eq. 38 implies that $\Xi = \mathbf{E} \times \dot{\mathbf{E}}$ is the quantity that couples to MCD. This quantity is maximized when $\mathbf{E}$ and $\dot{\mathbf{E}}$ are orthogonal, i.e. when the electric field describes a circle. The optical-frequency magnetic field and electric and magnetic gradients are irrelevant for MCD. Thus

MCD is maximized for circularly polarized light, and we do not expect to find sculpted fields with enhanced MCD. This finding contrasts with the enhancements predicted for chiral CD, and highlights the different physical origins of these two effects.

The expression for Ξ captures the physical picture of the electric field vector rotating about an axis, which is often associated with circular polarization. However, circular polarization is not necessary for **E** to describe a circle. Consider a Gaussian beam coming to a focus, with linear polarization in *x*, propagating in *z*. The electric field is given by the well-known expression[21].

$$E_x = E_0 \frac{w_0}{w(z)} \exp\left( \frac{-r^2}{w^2(z)} - ikz - ik\frac{r^2}{2z\left(1 + \left(\frac{z_R}{z}\right)^2\right)} + i\arctan\left(\frac{z}{z_R}\right) \right), \quad (39)$$

where $w_0$ is the beam waist, $z_R = \frac{\pi w_0^2}{\lambda}$ is the Rayleigh range, and $w(z) = w_0\sqrt{1 + \left(\frac{z}{z_R}\right)^2}$. There is also a component of the electric field along *z*. To first order in $\frac{\lambda}{w_0}$, $E_z$ is [22]:

$$E_z = \frac{i}{k}\frac{\partial E_x}{\partial x}. \quad (40)$$

Near the focus, the electric field rotates in the x-z plane, which generates a non-zero time average of $\mathbf{E} \times \dot{\mathbf{E}}$. At the beam waist, $z = 0$, we find:

$$\langle \mathbf{E} \times \dot{\mathbf{E}} \rangle_t = \left(0, \frac{-2\omega E_0^2 x}{k w_0^2} \exp\left(\frac{-2r^2}{w_0^2}\right), 0\right). \quad (41)$$

Therefore we predict that a DC magnetic field in the $\hat{y}$ direction gives rise to differential absorption across the beam cross section. This signal has the same origin as magnetic CD.

We propose an experiment to demonstrate this effect via fluorescence detected MCD (FDMCD), as illustrated in Figure 3. A thin film of fluorescent material is placed in the focus of a linearly polarized beam. An external magnetic field is applied transverse to the beam and the polarization. The small FDMCD signal is superimposed on the much larger E1-E1 fluorescence, so the FDMCD appears as a slight shift along the *x*-axis of the peak position of the fluorescence. By modulating the magnetic field, this peak is made to oscillate along *x*, leading to a fluorescence signal that can be detected with a spatially resolved detector and a lock-in amplifier.

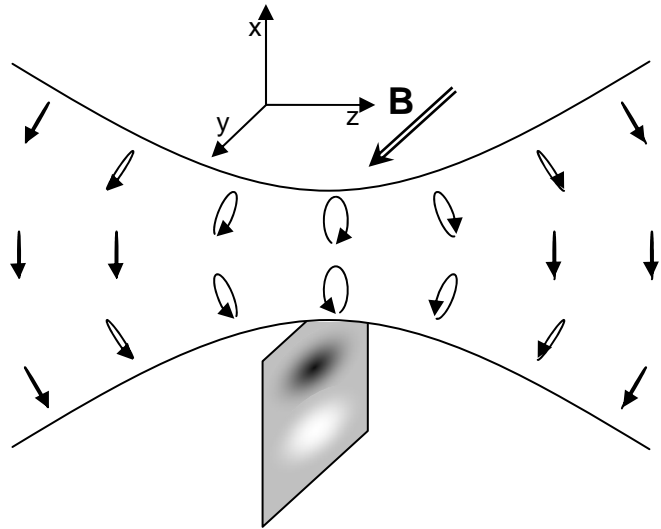

**Figure 3.** Magnetic circular dichroism in a focused linearly polarized beam. Near the focus, the electric field describes an ellipse in the *x-z* plane. A magnetic field along the *y* axis is predicted to induce a differential absorption across the beam.

**5. Relation to the Lipkin Zilch**

In 1964 Lipkin introduced ten new conserved electromagnetic quantities, but did not find a physical interpretation [14]. We recently showed that one of these terms, $Z^{000}$ is the same as our optical chirality, $C$ [13]. Here we show that the remaining nine terms are linear combinations of quantities that we derived above.

Consider six of Lipkin's terms given by the off-diagonal elements of

$$Z^{\alpha\beta 0} \equiv \delta_{\alpha\beta}[\mathbf{E}\cdot\nabla\times\mathbf{E} + \mathbf{H}\cdot\nabla\times\mathbf{H}] - E_\alpha(\nabla\times\mathbf{E})_\beta - E_\beta(\nabla\times\mathbf{E})_\alpha - H_\alpha(\nabla\times\mathbf{H})_\beta - H_\beta(\nabla\times\mathbf{H})_\alpha. \quad (42)$$

The three off-diagonal terms in $Z^{\alpha\beta 0}$ each separate into two terms of the form

$$-E_\alpha(\nabla\times\mathbf{E})_\beta - H_\beta(\nabla\times\mathbf{H})_\alpha. \quad (43)$$

Using Gaussian units with $c = 1$ as used in Lipkin's paper, and applying Maxwell's equations in free space, we have

$$E_\alpha \dot{B}_\beta - \dot{E}_\alpha B_\beta \rightarrow -E_\alpha(\nabla\times\mathbf{E})_\beta - H_\beta(\nabla\times\mathbf{H})_\alpha. \quad (44)$$

Therefore each off-diagonal element $Z^{\alpha\beta 0}$ is a linear combination of the EM quantities that drive E1-M1 transitions.

There are three additional conserved quantities from Lipkin of the form:

$$(\mathbf{E}\times\dot{\mathbf{E}} + \mathbf{H}\times\dot{\mathbf{H}})_\alpha. \quad (45)$$

These are components of the flux of optical chirality $C$, and are also related to MCD. We showed above that the quantity $\mathbf{E}\times\dot{\mathbf{E}}$ determines the extent of MCD. If we also consider the perturbation to the magnetic polarizability by a magnetic field, then an identical procedure would identify $\mathbf{H} \times \dot{\mathbf{H}}$ as the relevant quantity for magnetic dipole transitions. Magnetic dipole transitions are normally much weaker than electric dipole transitions, and can be neglected.

Conservation in vacuum for each of the Lipkin terms is a special case of a general conservation law for any quantity of the form:

$$Y = A\dot{B} - B\dot{A}, \qquad (46)$$

where $A$ and $B$ are components of $\mathbf{E}$ and $\mathbf{B}$ acted on by any linear operator, such as time or spatial derivative. We define the corresponding flux

$$\mathbf{\Phi} \equiv -A\nabla B + B\nabla A, \qquad (47)$$

and immediately obtain a conservation law in free space:

$$\dot{Y} + \nabla \cdot \mathbf{\Phi} = 0.$$

This procedure gives an infinite number of quantities that are conserved in vacuum. These quantities are likely to have the physical interpretation of being the couplings to molecular polarizability tensor components.

### 6. Discussion

We showed that there exist many configurations of the electromagnetic fields and field gradients that cannot be generated from a single plane wave; but that *any* local configuration of the field can be generated from a combination of plane waves. There exist achiral multipole transitions in oriented molecules that are completely invisible to plane waves, but that are detectable in suitably designed optical standing waves. Just as circular dichroism measurements probe aspects of molecular structure that are not detected by conventional absorption, we expect these achiral multipole transitions also to hold new molecular information. We also found that a tightly focused beam of linearly polarized light in a magnetic field probes the molecular properties usually probed by magnetic circular dichroism. Here we considered only linear optics, where the rate of excitation is quadratic in the fields. The field objects we considered here can appear in greater number when one considers nonlinear optics, and interesting effects may arise in that domain as well.[23]

**Appendix 1: Plane-wave basis for local field geometries**

At a single point in space, one requires sixteen independent complex quantities to describe the electric field, magnetic field, electric gradients and magnetic gradients for a time-harmonic solution to Maxwell's Equations. These quantities can be arranged in a sixteen-dimensional vector $\tilde{\mathbf{Q}}\exp(-i\omega t)$, where $\tilde{\mathbf{Q}}$ is complex. We choose to construct a basis for $\tilde{\mathbf{Q}}$ from plane waves of the form $\mathbf{E}^{(0)}\exp(i\mathbf{k}\cdot\mathbf{r}-i\omega t)\exp(i\phi)$, for which we can choose the phase $\phi$ to give the real and imaginary parts independently. Therefore it is sufficient to show that one can construct the real part of $\tilde{\mathbf{Q}}$ from plane waves. The imaginary part follows similarly.

In the following table, we list 16 plane waves and their associated vectors in the space of $\tilde{\mathbf{Q}}$, calculated at the origin $\mathbf{r} = 0$. A similar set with $\phi = \pi/2$ completes the basis for all possible EM fields and gradients at $\mathbf{r} = 0$. These vectors can be verified to be linearly independent by checking that the determinant of the 16 by 16 matrix is non-zero. Therefore these vectors form a basis for the vector space. Of course this choice of basis is not unique. We have chosen units of time and distance such that $k_0 = c = 1$. We only give five of the gradients of $\mathbf{E}$ and $\mathbf{B}$; the others are determined from Maxwell's Equations.

**TABLE 1.** Sixteen plane waves with phase $\phi = 0$, which span the real part of the space of possible electromagnetic fields and field gradients.

| k | E | B | $(E_{xy}, E_{yz}, E_{zx}, E_{xx}, E_{yy})$ | $(B_{xy}, B_{yz}, B_{zx}, B_{xx}, B_{yy})$ |
|---|---|---|---|---|
| $(0, 1, 0)$ | $(E_1, 0, 0)$ | $(0, 0, -E_1)$ | $(E_1, 0, 0, 0, 0)$ | $(0, 0, 0, 0, 0)$ |
| $(0, 0, 1)$ | $(E_2, 0, 0)$ | $(0, E_2, 0)$ | $(0, 0, 0, 0, 0)$ | $(0, E_2, 0, 0, 0)$ |
| $(1, 0, 0)$ | $(0, E_3, 0)$ | $(0, 0, E_3)$ | $(0, 0, 0, 0, 0)$ | $(0, 0, E_3, 0, 0)$ |
| $(0, 0, 1)$ | $(0, E_4, 0)$ | $(-E_4, 0, 0)$ | $(0, E_4, 0, 0, 0)$ | $(0, 0, 0, 0, 0)$ |
| $(1, 0, 0)$ | $(0, 0, E_5)$ | $(0, -E_5, 0)$ | $(0, 0, E_5, 0, 0)$ | $(0, 0, 0, 0, 0)$ |
| $(0, 1, 0)$ | $(0, 0, E_6)$ | $(E_6, 0, 0)$ | $(0, 0, 0, 0, 0)$ | $(E_6, 0, 0, 0, 0)$ |
| $(0, 0, -1)$ | $(E_7, 0, 0)$ | $(0, -E_7, 0)$ | $(0, 0, 0, 0, 0)$ | $(0, E_7, 0, 0, 0)$ |
| $(-1, 0, 0)$ | $(0, E_8, 0)$ | $(0, 0, -E_8)$ | $(0, 0, 0, 0, 0)$ | $(0, 0, E_8, 0, 0)$ |
| $(0, -1, 0)$ | $(0, 0, E_9)$ | $(-E_9, 0, 0)$ | $(0, 0, 0, 0, 0)$ | $(E_9, 0, 0, 0, 0)$ |
| $(\frac{1}{\sqrt{2}}, 0, -\frac{1}{\sqrt{2}})$ | $(E_{10}, 0, E_{10})$ | $(0, -\sqrt{2}E_{10}, 0)$ | $(0, 0, \frac{E_{10}}{\sqrt{2}}, \frac{E_{10}}{\sqrt{2}}, 0)$ | $(0, E_{10}, 0, 0, 0)$ |
| $(0, \frac{1}{\sqrt{2}}, -\frac{1}{\sqrt{2}})$ | $(0, E_{11}, E_{11})$ | $(\sqrt{2}E_{11}, 0, 0)$ | $(0, \frac{-E_{11}}{\sqrt{2}}, 0, 0, \frac{E_{11}}{\sqrt{2}})$ | $(E_{11}, 0, 0, 0, 0)$ |
| $(0, -1, 0)$ | $(E_{12}, 0, 0)$ | $(0, 0, E_{12})$ | $(-E_{12}, 0, 0, 0, 0)$ | $(0, 0, 0, 0, 0)$ |
| $(0, 0, -1)$ | $(0, E_{13}, 0)$ | $(E_{13}, 0, 0)$ | $(0, -E_{13}, 0, 0, 0)$ | $(0, 0, 0, 0, 0)$ |
| $(-1, 0, 0)$ | $(0, 0, E_{14})$ | $(0, E_5, 0)$ | $(0, 0, -E_5, 0, 0)$ | $(0, 0, 0, 0, 0)$ |
| $(\frac{1}{\sqrt{2}}, 0, -\frac{1}{\sqrt{2}})$ | $(0, E_{15}, 0)$ | $(\frac{E_{15}}{\sqrt{2}}, 0, \frac{E_{15}}{\sqrt{2}})$ | $(0, \frac{-E_{15}}{\sqrt{2}}, 0, 0, 0)$ | $(0, 0, \frac{E_{15}}{2}, \frac{E_{15}}{2}, 0)$ |
| $(0, \frac{1}{\sqrt{2}}, -\frac{1}{\sqrt{2}})$ | $(E_{16}, 0, 0)$ | $(0, \frac{-E_{16}}{\sqrt{2}}, \frac{-E_{16}}{\sqrt{2}})$ | $(\frac{E_{16}}{\sqrt{2}}, 0, 0, 0, 0)$ | $(0, \frac{E_{16}}{2}, 0, 0, \frac{-E_{16}}{2})$ |

**Acknowledgments**. This work was supported by an ONR Young Investigator Award and the DARPA Young Investigator Program. NY was partially supported by a fellowship from NSERC Canada.

References

(1) Krueger, B. P.; Scholes, G. D.; Fleming, G. R. *J. Phys. Chem. B* **1998**, *102*, 5378.

(2) Harbola, U.; Mukamel, S. *Phys. Rev. A* **2004**, *70*, 52506.

(3) Hachmann, J.; Dorando, J. J.; Aviles, M.; Chan, G. K. *J. Chem. Phys.* **2007**, *127*, 134309.

(4) Craig, D. P.; Thirunamachandran, T. *Molecular Quantum Electrodynamics;* Courier Dover Publications: New York, 1998.

(5) Akbar Salam, *Molecular Quantum Electrodynamics: Long-Range Intermolecular Interactions;* Wiley: USA, 2009.

(6) Moskovits, M. *Rev. Mod. Phys.* **1985**, *57*, 783.

(7) Nitzan, A.; Brus, L. E. *J. Chem. Phys.* **1981**, *75*, 2205.

(8) Liebermann, T.; Knoll, W. *Colloids Surf. A* **2000**, *171*, 115.

(9) Haynes, C. L.; Van Duyne, R. P. *J. Phys. Chem. B* **2003**, *107*, 7426.

(10) Sánchez, E. J.; Novotny, L.; Xie, X. S. *Phys. Rev. Lett.* **1999**, *82*, 4014.

(11) Sundaramurthy, A.; Schuck, P. J.; Conley, N. R.; Fromm, D. P.; Kino, G. S.; Moerner, W. E. *Nano Lett.* **2006**, *6*, 355.

(12) Cohen, A. E. *J. Phys. Chem. A* **2009**, *113*, 11084.

(13) Tang, Y.; Cohen, A. E. *Phys. Rev. Lett.* **2010**, *104*, 163901.

(14) Lipkin, D. M. *J. Math. Phys.* **1964**, *5*, 696.

(15) Candlin, D. J. *Il Nuovo Cimento* **1965**, *37*, 1390.

(16) Kibble, T. W. B. *J. Math. Phys.* **1965**, *6*, 1022.

(17) Barron, L. D. *Molecular Light Scattering and Optical Activity;* Cambridge University Press: Cambridge, 2004.

(18) Atkins, P. W.; Friedman, R. S. *Molecular quantum mechanics;* Oxford university press Oxford: 1983.

(19) Buckingham, A.; Dunn, M. *J. Chem. Soc. (A)* **1971**, *1971*, 1988.

(20) Healy, W. P. *J. Chem. Phys.* **1976**, *64*, 3111.

(21) Siegman, A. E. *Lasers;* University Science Books: Sausalito, CA, 1986.

(22) Davis, L. W. *Phys. Rev. A* **1979**, *19*, 1177.

(23) Mukamel, S. *Principles of Nonlinear Optical Spectroscopy;* Oxford University Press: USA, 1995.

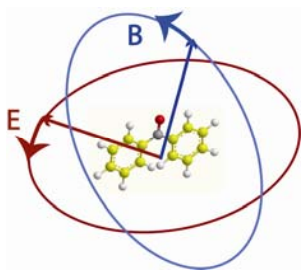